\documentclass{aastex6}
\usepackage{amssymb}
\usepackage{rotating}
\usepackage{graphicx}
\shorttitle{Harmonic summing: probabilities}
\shortauthors{M. Yu}
\begin{document}
\title{Harmonic summing improves pulsar detection sensitivity:
  probability analysis} \author{Meng Yu} \affil{National Astronomical
  Observatories, Chinese Academy of Sciences} \affil{CAS Key
  Laboratory of FAST, National Astronomical Observatories, Chinese
  Academy of Sciences, \\ 20A Datun Road, Chaoyang District, Beijing
  100101, P. R. China \\ E-mail: vela.yumeng@gmail.com}
\begin{abstract}
Practical application of the harmonic summing technique in the
power-spectrum analysis for searching pulsars has exhibited the
technique's effectiveness. In this paper, theoretical verification of
harmonic summing considering power's noise-signal probability
distribution is given. With the top-hat and the modified von
Mises pulse profile models, contours along which spectra total power
is expected to exceed the 3\,$\sigma$ detection threshold with 0.999
confidence corresponding to $m=1, 2, 4, 8, 16$, or 32 harmonics summed
are given with respect to the mean pulse amplitude and the pulse duty
cycle. Optimized numbers of harmonics summed relative to the duty
cycles are given. The routine presented builds a theoretical
estimate of the minimum detectable mean flux density,
i.e. sensitivity, under the power-spectrum searching method.
\end{abstract}
\keywords{pulsars:general --- stars:neutron}
\section{Introduction}\label{sec:intro}
Due to the sensitive response to periodicity, discrete Fourier
transform has widely been used in searching for pulsars. In this
technique, a dedispersed and possibly barycentered\footnote{For
  observations longer than typically 30\,min, Doppler shift in pulsar
  pulse frequency caused by motion of the earth appears to be
  significant; the barycentering process removes the shift as if the
  observation was carried out at the solar system barycenter, an
  approximate inertial reference system. Furthermore, time domain
  resampling or frequency domain correlation are often implemented to
  remove the Doppler shift caused by pulsar orbital motion as if the
  pulsar turns to be isolated with locating at the binary system
  barycenter.} $N$-point real time series $\mathcal{T}$ derived from
an observation is Fourier transformed into a complex spectrum series
$u_j + iv_j$ with point number $M_{\rm iid}^{(1)} = \frac{N}{2}+1$,
where $i$ is the unit of imaginary number and $j$ is the number index
ranging from 0 to $M_{\rm iid}^{(1)} - 1$. Power spectrum is
subsequently formed as $w_j^{(1)} = u_j^2 + v_j^2$. For a Gaussian
white noise series $\mathcal{T}_{\rm noise}$, the derived $u_j$'s,
$v_j$'s and $w_j^{(1)}$'s are identical and independent (iid) random
variables respectively; central limit theorem expects any sample in
the $u_j$'s or $v_j$'s is Gaussian distributed and so, any sample in
the $w_j^{(1)}$'s is $\chi_2^2$ distributed or $\chi^2$ distributed
with 2 degrees of freedom. Average and variance of the $w_j^{(1)}$'s
are the variance of the noise series $\mathcal{T}_{\rm noise}$
multiplied by the point number $N$, $N{\rm Var}(\mathcal{T}_{\rm
  noise})$ \citep{gro75d,rem02}. Thus when normalizing the
$w_j^{(1)}$'s via dividing by $N{\rm Var}(\mathcal{T}_{\rm noise})$,
any sample in the resultant $\hat{w}_j^{(1)}$'s $\chi_2^2$ distributes
with unity average and variance. The probability for any
$\hat{w}_j^{(1)}$ sample to exceed some power $P$ is
$\mathbb{P}(\hat{w}_j^{(1)} > P) = e^{-P}$. Then the probability for
all $\hat{w}_j^{(1)}$'s to be smaller or equal to the power $P$ is
$\mathbb{P}(\hat{w}_j^{(1)}$'s $\leqslant P) = (1 - e^{-P})^{M_{\rm
    iid}^{(1)}}$. When letting the probability be the confidence level
$\mathbb{C}_{3\sigma} \sim 0.999$ (the value when integrating the
standard Gaussian distribution probability density from $-\infty$ to
$+3$), the power $P_{3\sigma}^{(1)}$ (see Eq. 14 in
\citealt{vvw+94}\footnote{The normalizing power used in \citet{vvw+94}
  is two times smaller than that used here, or powers there are two
  times larger.}) derived is the 3\,$\sigma$ detection threshold. Any
$\hat{w}_j^{(1)}$ sample that exceeds the power should be noticed as
the probability for this to be induced by noise is only
$1-\mathbb{C}_{3\sigma} = 0.001$; signal is much more likely to have
presented.

The $\hat{w}_j^{(1)}$'s can be summed with each other. For sum with
$m=2$, in the ``Lyne-Ashworth'' routine for example, one stretches the
original spectrum by a factor of two by repeating each
$\hat{w}_j^{(1)}$'s once as the next sample, then adds the
intermediate series to the original series to form the power spectrum
$\hat{w}_j^{(2)}$. In this spectrum, samples $\chi_4^2$ distribute and
the mean and the variance of the spectrum are both 2. As the summation
is implemented between two iid samples, number of iid samples is
reduced to $M_{\rm iid}^{(2)} = \frac{M_{\rm iid}^{(1)}}{2}$ (though
length of the spectrum is still $M_{\rm iid}^{(1)}$). The summation is
typically implemented three more times with $m=4, 8$ and
16\footnote{In the ``Lyne-Ashworth'' routine for $m=4$ summation, the
  intermediate series is not formed by simply repeating the
  $\hat{w}_j^{(2)}$ samples and so does the summation with $m=8$ or
  16. Please refer to the pulsar search software package
  \textsc{sigproc} for the source code.}. In the resultant
$\hat{w}_j^{(m)}$'s, samples $\chi_{2m}^2$ distribute with the iid
sample number $M_{\rm iid}^{(m)} = \frac{M_{\rm
    iid}^{(1)}}{m}$. Average and variance of the $\hat{w}_j^{(m)}$'s
are $m$. The 3\,$\sigma$ detection threshold $P_{3\sigma}^{(m)}$ can
be derived numerically as how $P_{3\sigma}^{(1)}$ is derived with
considering $\mathbb{P}(\hat{w}_j^{(m)} > P) = \sum_{k=0}^{m-1}
\frac{P^k}{k!}e^{-P}$ which is the probability for any
$\hat{w}_j^{(m)}$ sample to exceed the power $P$. These summing
processes have been named ``harmonic summing'', because, by
implementing the summation, powers at the fundamental and harmonic
frequencies of a supposed signal can be added. In the sum with $m=2$,
powers at the signal's fundamental and 2nd harmonic frequencies are
summed at the signal's 2nd harmonic frequency. In the sum with $m=4$,
powers at the signal's fundamental up to 4th harmonic frequencies are
summed at the signal's 4th harmonic frequency. Results of the sum with
higher $m$ are similar. As complex phase has been lost when forming
the power spectra, harmonic summing is incoherent summation.

In the power spectrum of a time series containing noise and signal
simultaneously, the linear attribute of Fourier transform makes both
the real and imaginary parts of sample $j$ where the signal is are sum
of the noise and signal components, or $u_{{\rm tot},j} = u_{{\rm
    noise},j} + u_{{\rm sig},j}$, $v_{{\rm tot},j} = v_{{\rm noise},j}
+ v_{{\rm sig},j}$.  Then the total power $w_{{\rm tot},j}^{(1)}$ and
the signal power $w_{{\rm sig},j}^{(1)}$ are $w_{{\rm tot},j}^{(1)} =
u_{{\rm tot},j}^2 + v_{{\rm tot},j}^2$ and $w_{{\rm sig},j}^{(1)} =
u_{{\rm sig},j}^2 + v_{{\rm sig},j}^2$ respectively. Because the noise
and signal are summed coherently, distribution of the normalized total
power $\hat{w}_{{\rm tot},j}^{(m)}$ is not the noise's $\chi_{2m}^2$
distribution shifted by the constant signal power $\hat{w}_{{\rm
    sig},j}^{(m)}$ \citep{vvw+94}, but follows the two-dimensional
noise-signal distribution with probability being from 0 to some power
$\hat{w}_{{\rm tot},j}^{(m)}$ determined by the cumulative probability
distribution function $\mathbb{F}_m (\hat{w}_{{\rm
    tot},j}^{(m)};\hat{w}_{{\rm sig},j}^{(m)})$ (see Eq. 16 in
\citealt{gro75d} or Eq. 19 in \citealt{vvw+94}). Consequently, $1 -
\mathbb{F}_m (P_{3\sigma}^{(m)};P_{{\rm sig},3\sigma}^{(m)}) = 0.999$
derives the $P_{{\rm sig},3\sigma}^{(m)}$, given which total power
$\hat{w}_{{\rm tot},j}^{(m)}$ is expected to exceed the detection
threshold $P_{3\sigma}^{(m)}$ with probability 0.999; \citet{vvw+94}
have provided numerical routines for realizing this. As a discrete
$N$-point sinusoid series with amplitude $a$ establishes signal power
$\left(\frac{1}{2}Na\right)^2$ \citep[see Eqs. 15 and 16 in][]{rem02},
minimum detectable pulse amplitude, i.e. sensitivity, at the
3\,$\sigma$ confidence level can be derived as long as the relation
between the $P_{{\rm sig},3\sigma}^{(m)}$ and a pulse profile model is
established \citep{vvw+94}. We see cases of two profile models next.
\section{The contours}\label{sec:amp}
\begin{figure}
\begin{center}
\begin{tabular}{r}
\includegraphics[width=9.67cm,angle=-90]{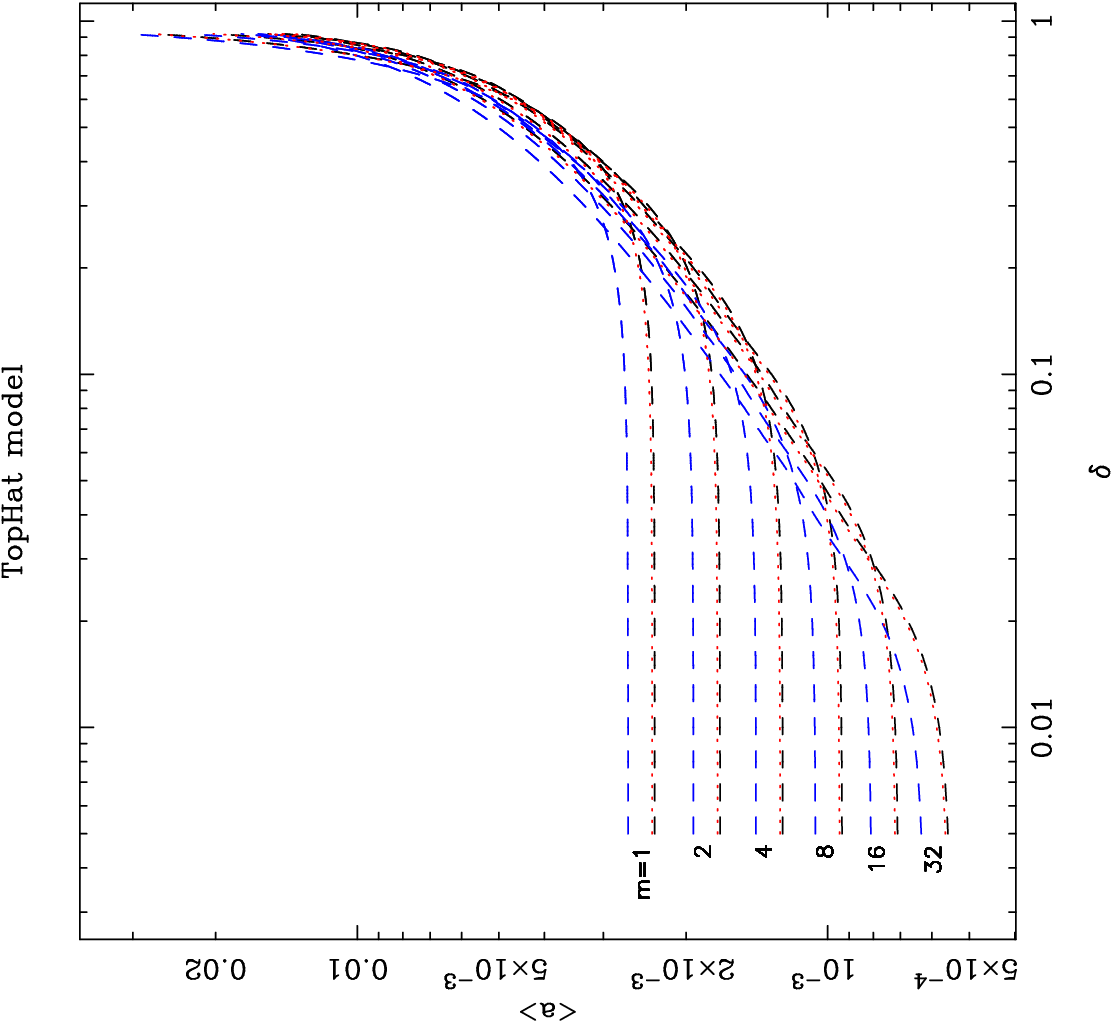} \\
\includegraphics[width=9cm,angle=-90]{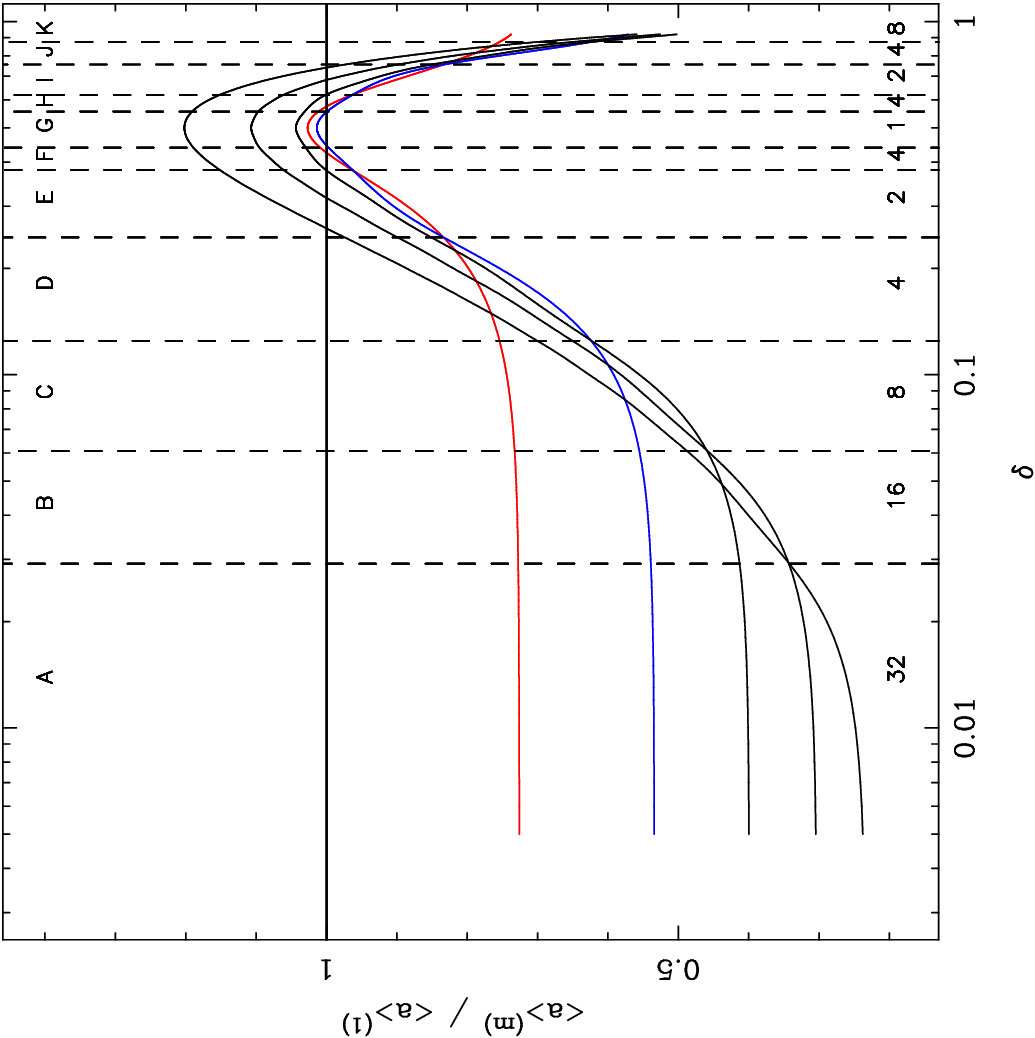}
\end{tabular}
\end{center}
\caption{Upper panel: The contours as a function of the pulse duty
  cycle ($x$-axis) and the mean pulse amplitude ($y$-axis) under the
  top-hat pulse profile model. Along the contours, powers of signal
  with $m = 1, 2, 4, 8, 16$ or 32 harmonics summed enable total powers
  at the signal frequency to exceed the 3\,$\sigma$ detection
  thresholds at confidence level 0.999. The black dashed, blue dashed
  and red dotted lines respectively indicate the integer frequency
  case, the fractional frequency case and the fractional frequency
  case with interbinning interpolation implemented. Note the contours
  were derived with a $2^{23}$ point Gaussian white noise
  series. Lower panel: The contours alternatively plotted with the
  $y$-axis changed into the amplitudes relative to the value derived
  without harmonic summing. The numbers at the bottom are the optimum
  numbers of harmonics summed. The letters at the top label the
  various duty cycle intervals whose boundaries are indicated by the
  vertical dashed lines.}\label{fig:ps_ct}
\end{figure}

One is the top-hat profile model. The top-hat or rectangular function
is described by an amplitude $a$ which is the difference between the
higher and lower levels, and the duty cycle $\delta$ which is the
ratio of span of the higher level to the domain of the function. For a
continuous periodic top-hat function, Fourier coefficient of $m$th
harmonic is $2a\delta{\rm sinc}(m\pi\delta)$. Then, in a discrete
$N$-point periodic top-hat series, $m$th harmonic establishes power
$\left[Na\delta{\rm sinc}(m\pi\delta)\right]^2$ in power spectrum
$w_j^{(1)}$. Thus, the minimum detectable amplitude $a$ for duty cycle
$\delta$ can be derived with
\begin{equation}
\sum_{k=1}^m \left[Na\delta{\rm sinc}(k\pi\delta)\right]^2 = P_{{\rm
    sig},3\sigma}^{(m)}P_{\rm norm}, \label{eq:p2a}
\end{equation}
where $P_{\rm norm}$ is the normalization power. An experiment was
done to realize the derivation. Firstly, a $2^{23}$ point white noise
series was generated, each sample was drawn from the standard Gaussian
distribution. Secondly, the series was Fourier transformed with
forming power spectrum $w_j^{(1)}$ ($j = 0, 1, \ldots,
2^{22}+1$). Thirdly, the $w_j^{(1)}$'s were normalized with their
average \citep{rem02}. Fourthly, the ``Lyne-Ashworth'' routine was
implemented for harmonic summing with obtaining $P_{3\sigma}^{(1)} =
21.8570$, $P_{{\rm sig},3\sigma}^{(1)} = 43.5297$, $P_{3\sigma}^{(2)}
= 24.3985$, $P_{{\rm sig},3\sigma}^{(2)} = 45.9351$,
$P_{3\sigma}^{(4)} = 28.8732$, $P_{{\rm sig},3\sigma}^{(4)} =
49.7991$, $P_{3\sigma}^{(8)} = 36.6707$, $P_{{\rm sig},3\sigma}^{(8)}
= 55.7288$, $P_{3\sigma}^{(16)} = 50.2974$, $P_{{\rm
    sig},3\sigma}^{(16)} = 64.5139$ and $P_{3\sigma}^{(32)} =
74.4269$, $P_{{\rm sig},3\sigma}^{(32)} = 77.2078$. Distributions of
the $\hat{w}_j^{(m)}$'s were found to be in agreement with the
distributions in theory. Finally, detectable minimum mean amplitudes
$\langle a \rangle = a\delta$ at the 3\,$\sigma$ confidence level for
$m = 1, 2, 4, 8, 16$ and $32$ with respect to $\delta \in [0.005,
  0.92]$ were derived. Note, in the experiment, the ``Lyne-Ashworth''
routine was extended with $m$ up to 32. The results are shown by the
black dashed lines in Fig. \ref{fig:ps_ct} upper panel. Ratio
  $\frac{\langle a \rangle^{(m)}}{\langle a \rangle^{(1)}}$ is shown
  in the lower panel. In Table \ref{tab:nhrm}, the optimum numbers of
  harmonics summed and the corresponding duty cycle intervals are
  given. Because of the equivalence of the wide pulses to narrow
  negative pulses, the optimum numbers of harmonics summed and the
  duty cycle intervals are symmetric relative to the 0.5 duty cycle.

\begin{figure}
\begin{center}
\begin{tabular}{r}
\includegraphics[width=9.67cm,angle=-90]{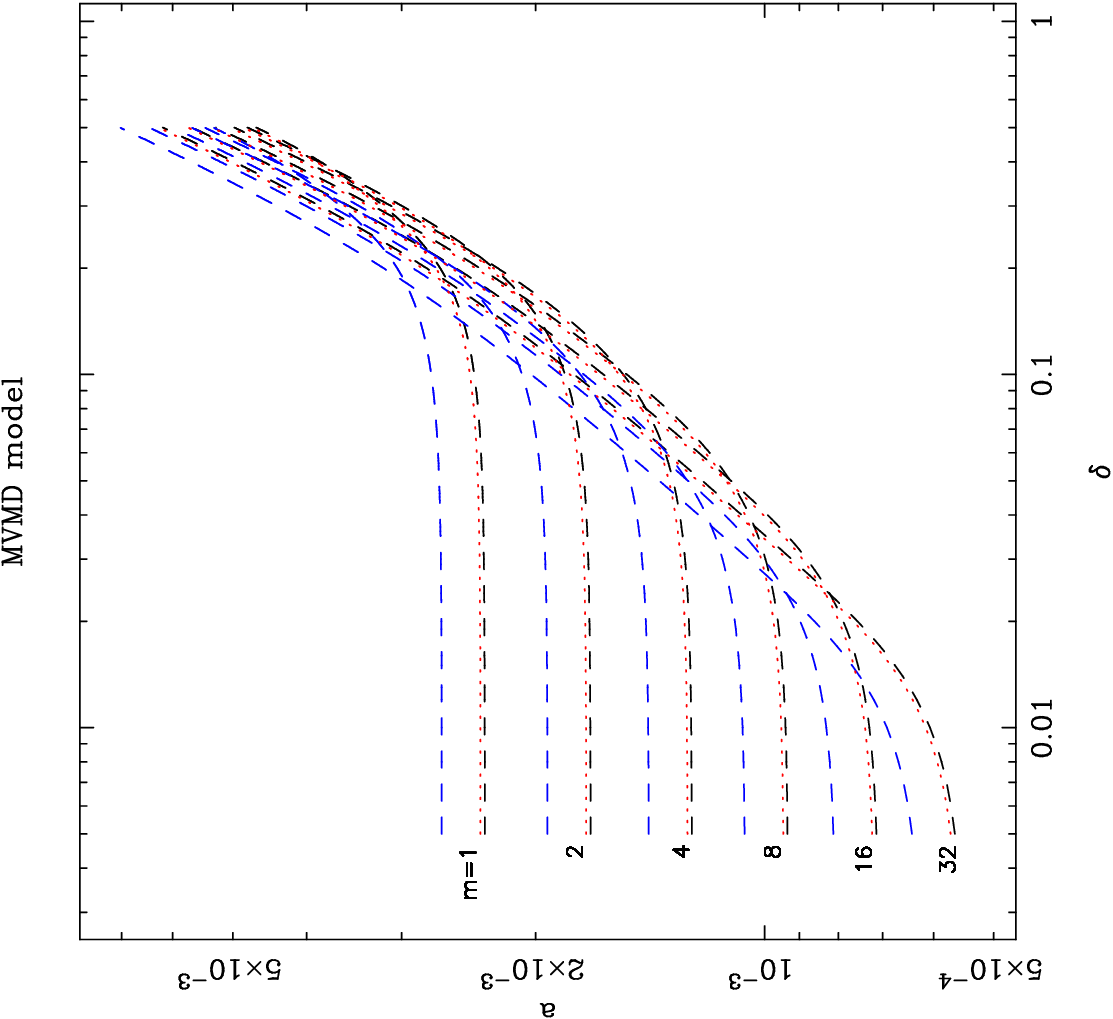} \\ 
\includegraphics[width=9cm,angle=-90]{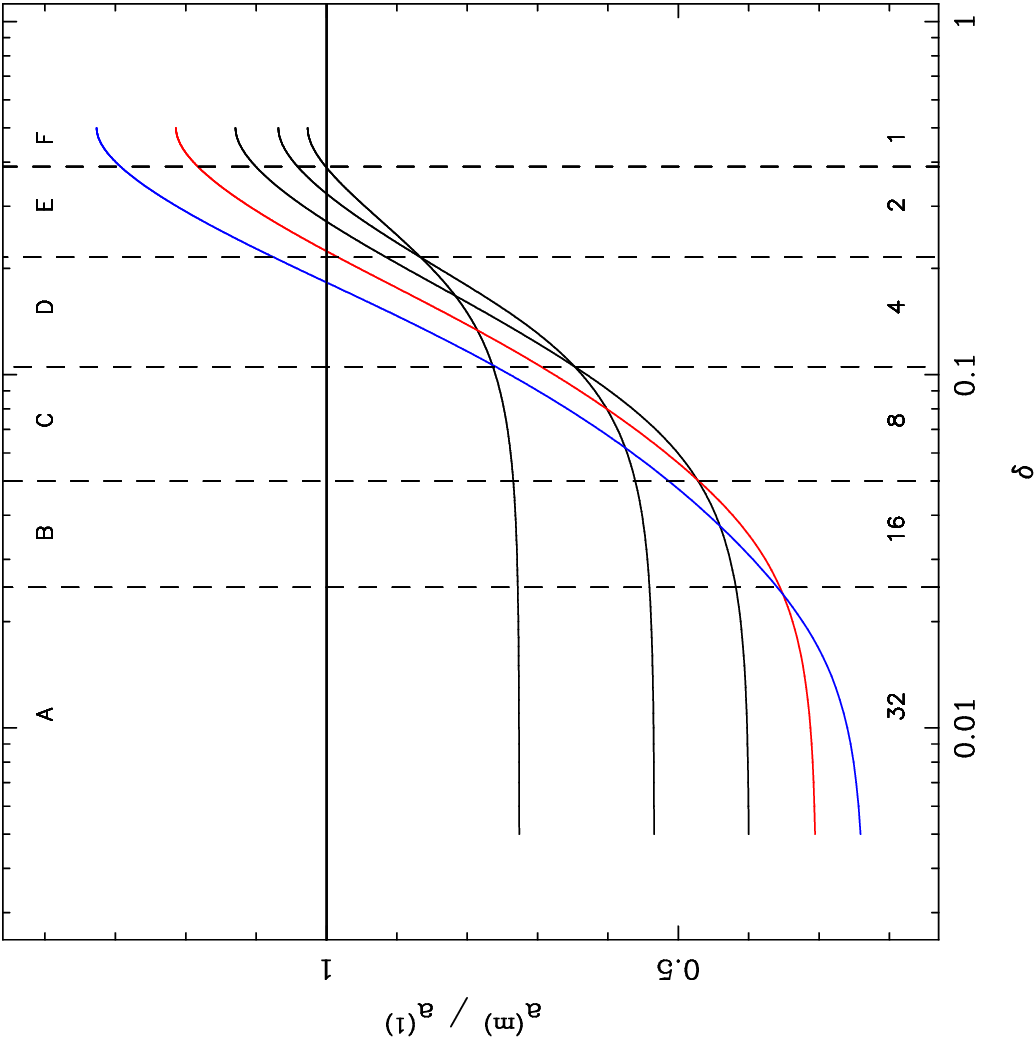}
\end{tabular}
\end{center}
\caption{The contours derived under the modified von Mises pulse
  profile model. The interpretations of the plots are the same as
  those in Fig. \ref{fig:ps_ct}.}\label{fig:ps_ct_vm}
\end{figure}

The other model is the modified von Mises profile model (MVMD)
\citep[see Eq. 20 in][]{rem02}. For this model, the equivalent width,
which is the division between the area under the function (the $a$)
and the function's maximum \citep[Eq. 22 in][]{rem02}\footnote{Erratum
  of Eq. (21) and Eq. (22) in \citet{rem02}: the full width at
  half-maximum (FWHM) of the modified von Mises distribution should be
  $\pi^{-1}\,{\rm arccos}\left[\frac{{\rm ln}\left({\rm
        cosh}\,\kappa\right)}{\kappa}\right]$; the maximum of the
  distribution should be $\frac{2a{\rm sinh}\,\kappa}{I_0(\kappa) -
    e^{-\kappa}}$.}, is $w_{\rm e} = \frac{I_0 (\kappa) -
  e^{-\kappa}}{2{\rm sinh}\,\kappa}$. In subsequent analysis regarding
this model, this $w_{\rm e}$ is used to define the pulse duty cycle
$\delta$. For pulse phase in pulsar rotation, $\delta = w_{\rm e}$ and
the $a$ is the mean pulse amplitude. For a continuous periodic MVMD
function, Fourier coefficient of the $m$th harmonic is $\frac{2a\,I_m
  (\kappa)}{I_0 (\kappa) - e^{-\kappa}}$. So, in a discrete $N$-point
MVMD series, the $m$th harmonic establishes power $\left[\frac{Na\,I_m
    (\kappa)}{I_0 (\kappa) - e^{-\kappa}}\right]^2$ in power spectrum
$w_j^{(1)}$. When using the power to replace the $\left[Na\delta{\rm
    sinc}(m\pi\delta)\right]^2$ part in Eq. \ref{eq:p2a} with
implementing the same experiment as for the top-hat profile model, the
minimum detectable mean pulse amplitude was derived. In the
computation, to obtain the concentration parameter $\kappa$
corresponding to a specific $\delta$, the bisection method was used to
find the root of the equation $\frac{I_0 (\kappa) - e^{-\kappa}}{2{\rm
    sinh}\,\kappa} - \delta = 0$. Because $\kappa$ increases
dramatically as $\delta$ becomes smaller, the equation could only be
solved for the $\delta$ larger than 0.03. For $\delta < 0.03$, the
$\kappa$ values were calculated as $\frac{1}{2\pi\delta^2}$, since the
modified Bessel function $I_m (\kappa)$ approaches
$\frac{e^\kappa}{\sqrt{2\pi\kappa}}$ when $\kappa \to +\infty$. In the
large $\kappa$ limit, the exponentially scaled modified Bessel
function is used to approximate the ratio of $I_m (\kappa)$ to $I_0
(\kappa)$, i.e.  $\frac{I_m (\kappa)}{I_0 (\kappa)} \sim
\frac{e^{-\kappa}I_m (\kappa)}{e^{-\kappa}I_0 (\kappa)}$. The derived
minimum detectable mean amplitudes are shown as the black dashed line
in Fig. \ref{fig:ps_ct_vm} upper panel. In the lower panel, ratio
$\frac{a^{(m)}}{a^{(1)}}$ is shown. The optimum numbers of harmonics
summed and the corresponding duty cycle intervals are given in Table
\ref{tab:nhrm}.

The analysis above is restricted to integer frequencies, i.e.
integers between 1 and $M_{\rm iid}^{(1)}-1$. This refers to the case
when power spectrum happens to sample the frequency of a
signal. Signals having fractional frequencies are the more general
cases, in which the ``scalloping effect'' occurs \citep{rem02}. In the
power spectrum derived from a $N$-point sinusoid series with amplitude
$a$, power at the nearest integer frequency with difference $\Delta
\in [0.5,0.5]$ away from the signal frequency is the multiplication
between the power $\left(\frac{1}{2}Na\right)^2$ and the factor ${\rm
  sinc}^2 (\pi\Delta)$ \citep{rem02}. For narrow pulse cases, harmonic
summing algorithms in principle call the spectra with frequencies
closest to the frequencies of the signal's harmonics\footnote{In
  practice, a test for the ``Lyne-Ashworth'' algorithm with a
  fractional signal frequency showed, out of the 38 frequency bins
  called, 22 had an absolute offset (difference between the integer
  frequency called and the frequency of the specific harmonic) $<0.5$,
  13 had an absolute offset between 0.5 and 1.0, and 3 had an absolute
  offset $>1.0$; the largest offset is $+1.254$. Note same test on
  different harmonic summing algorithms would lead to different
  results, so it is necessary to include the performances when
  building a realistic sensitivity estimate for a specific searching
  program.}. So on average the scalloping effect causes a 23 per cent
loss of the signal power \citep{klis89,vvw+94} and an efficient factor
$\gamma = 0.77$ can be multiplied to the left hand side of
Eq. \ref{eq:p2a} to take the effect into account. The derived contours
on the $\delta - a$ plane under the 3\,$\sigma$ confidence level are
presented by the blue dashed lines in Figs. \ref{fig:ps_ct},
\ref{fig:ps_ct_vm} upper panels for the top-hat and MVMD profile
models, respectively. Relative amplitudes are the same as those of the
integer frequency case since the $\gamma$ factor is a constant.

An effective method to overcome the scalloping effect is the Fourier
interpolation. In this method, complex spectra with frequency locating
at any position between adjacent two integer frequencies is formed as
the weighted coherent sum of the spectra at $m$ integer frequencies
around \citep[see Eq. 30 in][]{rem02}. Since large $m$ leads to
expensive computation, the ``interbinning'' case is popular. This
corresponds to the $m=2$ Fourier interpolation but changes the
coefficient from $\frac{2}{\pi}$ to $\frac{\pi}{4}$ to boost the
response at half-integer frequency to be the full response \citep[see
  Eq. 31 in][]{rem02}. The interbinning interpolation raises the
efficient coefficient to $\gamma = 0.97$ on average. The derived
contours are then presented as the red dotted lines in the upper
panels in Figs. \ref{fig:ps_ct} and \ref{fig:ps_ct_vm}.
\section{Discussion}\label{sec:disn}
To determine the detectable minimum mean flux density or sensitivity
is an essential requirement of a pulsar search program. This is a
complicated problem. As described in \citet{cc97}, sensitivity is a
function of the radiometer noise, intrinsic pulse profile, pulsar
period and dispersion measure (DM), and the method used to find
pulsars. Level of the radiometer noise or rms fluctuation in system
temperature $T_{\rm sys}$ is, as manifested by the radiometer equation
\citep[see e.g. Eq. 12 in ][]{oneil02}, proportional to the $T_{\rm
  sys}$ itself. $T_{\rm sys}$ is a function of the source position,
the telescope pointing and the observing frequency; its sophisticated
calibration procedures were described by \citet{oneil02}. It has
usually been found observed time series exhibits the ``red'' power
spectral features. There are both natural and artificial sources that
induce the red noises. The natural sources are, for example, the
emission from background and/or foreground celestial bodies
\citep{is96} and the variations of the atmospheric emission
\citep{oneil02}. The artificial sources are more diverse. For example,
the dependence of the temperature from the ground on the telescope
azimuth, zenith angles, the dependence of temperature from the
atmosphere on the telescope zenith angle, the instability of the
receiving system and the dependence of antenna gain on telescope
elevation \citep{oneil02}.

Another strong artificial source is the radio frequency interference
(RFI). The RFI is more complicated in that apart from it is telescope
dependent it varies from time to time. Although multiple efforts,
including active surface and hardware/software filters, have been made
for removing the red noises, they cannot be eliminated completely. By
simulating pulsar signals in real observations, \citet{lbh+15}
incorporated RFI into the analysis of sensitivity for the PALFA
survey. They found at the long period end the predicted sensitivities
were degraded by a factor of $\sim$3 to $\sim$7 compared to the
predictions made with the \citet{dtws85} method. \citet{pkr+18} have
further analyzed PALFA sensitivities for long period pulsars; similar
results were obtained.

The approach implemented by \citet{dtws85} is to examine the
significance of an averaged top-hat pulse profile out of a given flux
density. This is realized by first applying the radiometer equation to
the top-hat pulse signal (see Eq. 1 in \citealt{dtws85} or the
Appendix A1.4 in \citealt{lk12} for the detailed derivation), then
setting the entire integration time per telescope pointing as the
observing integration time in the equation. The significance is
indicated by the signal-to-noise ratio which is defined as the
proportion of height of the top-hat to the rms radiometer noise and is
statistically modeled by Gaussian distributions \citep[see \S7.1.1.1
  in ][]{lk12}. With the integration time per pointing 2.3\,min and
the sampling time 16.8\,ms, \citet{dtws85} set the confidence limit
$\sim$7.5\,$\sigma$ under the profile signal-to-noise. Because the
detection sensitivity is partially a function of the searching method
\citep{cc97}, the approach described is not appropriate since the
Fourier domain method was used by \citet{dtws85} to implement their
search. The detection confidence limit should be given under the
statistics in the Fourier domain rather than the statistics of pulse
profile. The threshold (profile height) implied out of the profile
signal-to-noise is not consistent with the threshold (spectra power)
implied in the Fourier domain. The \citet{dtws85} method has
subsequently been implemented by \citet{jlm+92}, \citet{mlc+01} and
\citet{cfl+06} for their respective surveys, though the Fourier method
was also used for searching pulsars. For the Parkes multi-beam pulsar
survey, \citet{cra00} and \citet{mlc+01} implemented a semi-analytic
approach to obtain the sensitivity estimates. But that was for
including the harmonic summing into the \citet{dtws85} method; the
inconsistence issue remains.

\citet{vvw+94} have proposed the approach to give sensitivity estimate
for the Fourier domain searching method in the power-spectrum
manner. They implemented their method with the sinusoidal pulse
profile for the X-ray pulsar search. In radio pulsar search where
narrow pulses are more commonly seen, the relations between the
spectra thresholds and the top-hat and MVMD profile models have been
presented in this work. The conversion of the derived minimum
detectable mean amplitudes (shown in Figs. \ref{fig:ps_ct} and
\ref{fig:ps_ct_vm} upper panels) into the sensitivity values in the
unit of Jansky would be complicated, because a realistic conversion
should include the calibration of system temperature, the response of
bandpass and the RFI etc.; these are telescope dependent. However, out
of the purpose of illustrating the idea of the conversion, we see an
example below. In brief, the radiometer equation will be used for
individual pulses since the amplitudes derived are the values in one
pulse period. In the example, published system parameters of the
Parkes multi-beam pulsar survey are used.

\begin{figure}
\begin{center}
\begin{tabular}{c}
\includegraphics[width=9cm,angle=-90]{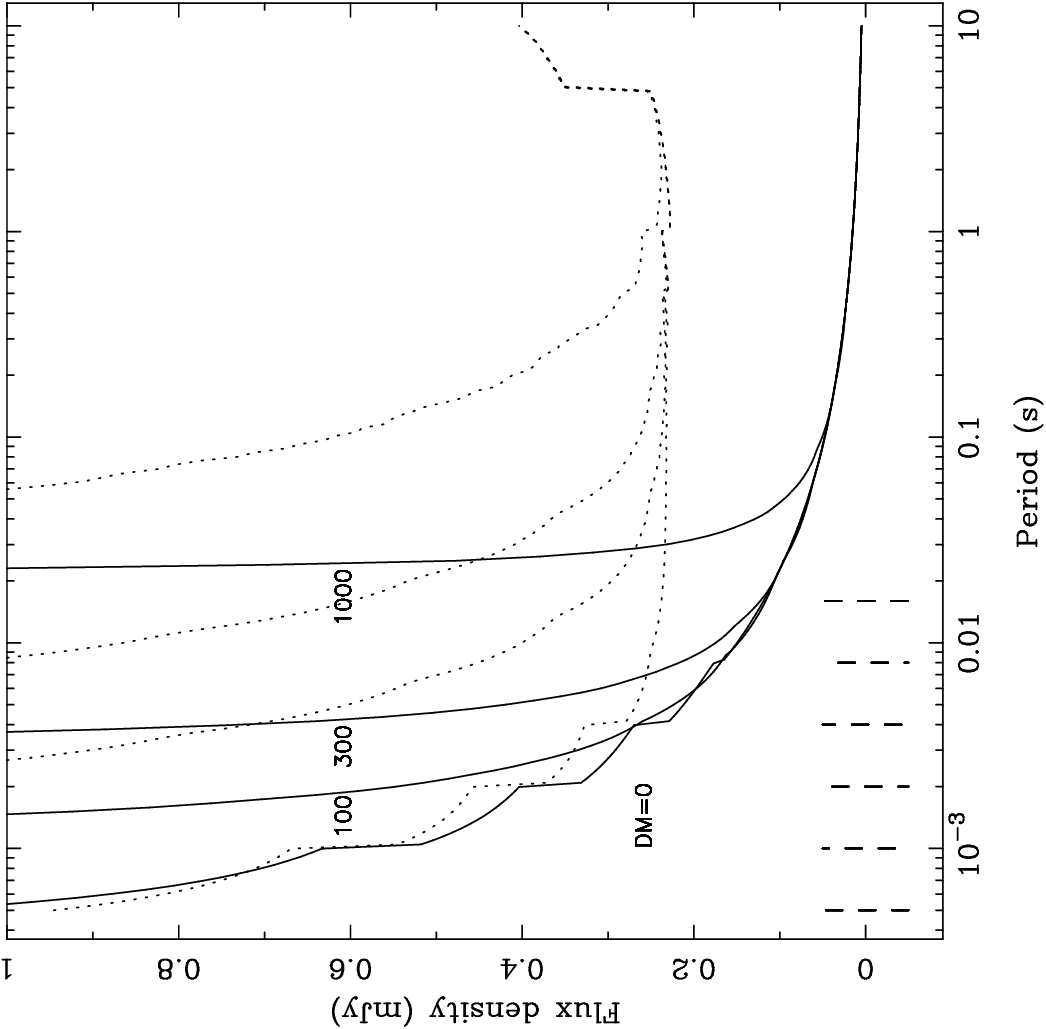} \\
\includegraphics[width=9cm,angle=-90]{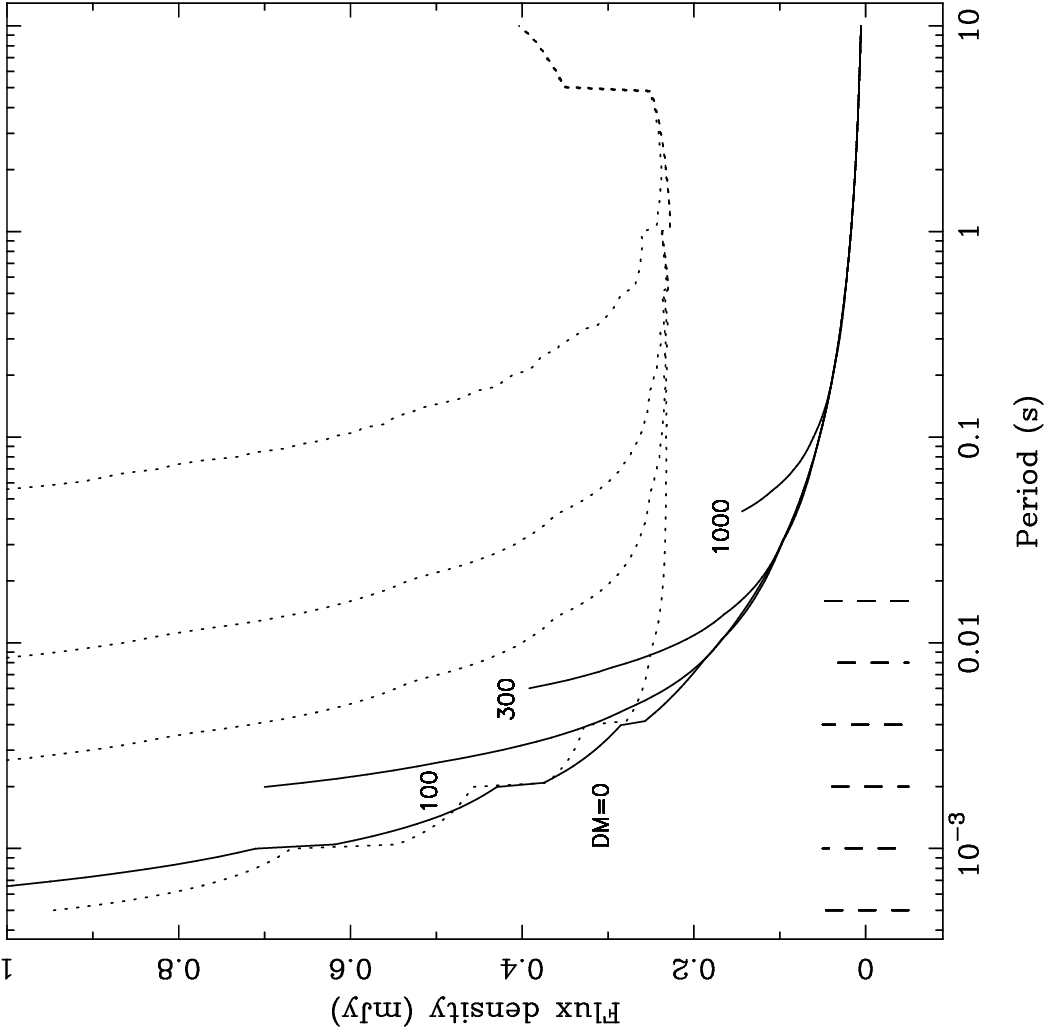}
\end{tabular}
\end{center}
\caption{The minimum detectable mean flux densities at DM = 0, 100,
  300 and 1000\,cm$^{-3}$\,pc for pulsar periods from 0.50\,ms to
  10.0\,s. The values were derived with the published system
  parameters of the Parkes multi-beam pulsar survey under the
  3\,$\sigma$ confidence limit. The upper and lower panels
  respectively correspond to the top-hat and the modified von Mises
  profile models. In the calculation, a 4 per cent pulse intrinsic
  duty cycle was assumed. The vertical dashed lines at the corner,
  from left to right, indicate periods 1.0, 2.0, 4.0, 8.0 and
  16.0\,ms, above which harmonic $m = 2, 4, 8, 16$ or 32 begins to
  present in the power spectrum. The dotted lines represent the
  sensitivities derived with the routine which is for the original
  sensitivity estimates of the multi-beam survey. In the calculation,
  a 4 per cent pulse intrinsic duty cycle and a 8\,$\sigma$ confidence
  limit were also assumed.}\label{fig:flux}
\end{figure}

In the multi-beam survey, the sampling interval $t_{\rm samp}$ was
configured as 0.25\,ms. Then the 35\,min observation produced $2^{23}$
samples in the time series. The shortest pulsar period the survey
detects is 0.50\,ms. At this period, only the fundamental presents in
the power spectrum (no harmonic presents). At zero DM, the effective
pulse width $W_{\rm e}$ is the quadrature sum of intrinsic pulse width
$W_0$ and the sampling interval. When assuming the intrinsic pulse
width to be 0.04 of the pulsar period, we have $W_{\rm e} = \left[
  (0.04 \times 0.5)^2 + 0.25^2 \right]^{\frac{1}{2}} \sim
0.25$\,ms. Duty cycle is then $\sim$0.50. Since no harmonic presents
at the 0.50\,ms period, among the minimum mean amplitudes
corresponding to this duty cycle (see Figs. \ref{fig:ps_ct} and
\ref{fig:ps_ct_vm} upper panels), the $m=1$ amplitudes should be taken
to calculate the profile signal-to-noise. In the integer frequency
case, the amplitude derived with the top-hat profile model was
$\sim$0.0072 while the amplitude derived with the MVMD profile model
was $\sim$0.0091. As standard deviations of the simulated white noise
series were derived as $\sim$1.0 for both of the realizations for the
profile models, the signal-to-noise values are then $\sim$0.0072 and
$\sim$0.0091 respectively. For the other system parameters, the survey
configured the antenna gain $G = 0.735$\,K\,Jy$^{-1}$, the
polarization number $n_{\rm pol} = 2$, the central frequency $f_{\rm
  ctr} = 1,374$\,MHz, the bandwidth $\Delta f_{\rm bw} = 288$\,MHz,
the digitization loss factor $\beta = 1.5$ and the receiver
temperature $T_{\rm rcvr} = 21$\,K \citep{mlc+01}. For the sky
temperature $T_{\rm sky}$, it is set 427\,K; this is the value of the
sky position with Galactic longitude 350.019$^\circ$ and Galactic
lattitude $-$0.677$^\circ$ measured at 408\,MHz
\citep{hssw82}\footnote{The temperature value is from the data with no
  filtering downloaded from
  https://lambda.gsfc.nasa.gov/product/foreground/fg\_haslam\_get.cfm.}. With
the average spectral index $-$2.5 of the sky background
\citep{hssw82}, we have 19\,K as the position's temperature at the
central frequency. With neglecting all other contributions to $T_{\rm
  sys}$, we have $T_{\rm sys} = T_{\rm rcvr} + T_{\rm sky} \sim
40$\,K. Thus, sensitivity at the 0.50\,ms period and zero DM was
derived as $\sim$1.1\,mJy under the top-hat profile model or as
$\sim$1.3\,mJy under the MVMD profile model. In the calculation, the
integration time in the radiometer equation was taken as the pulsar
period. The routine described above can be used up till period
1.0\,ms. From 1.0 to 2.0\,ms, with the emergence of the second
harmonic, the $m=2$ amplitudes should be taken if they are smaller
than the $m=1$ amplitudes. From 2.0 to 4.0\,ms $m=4$ amplitudes are
preferred, from 4.0 to 8.0\,ms $m=8$ amplitudes are preferred, from
8.0 to 16.0\,ms $m=16$ amplitudes are preferred and from 16.0\,ms on,
$m=32$ amplitudes are preferred. The derived sensitivities for pulsar
periods from 0.50\,ms to 10.0\,s and DM zero are shown in
Fig. \ref{fig:flux}.

For non-zero DM, the quadrature sum for the effective pulse width
should additionally include the pulse smearing times induced by
dispersion $t_{\rm DM}$ and scattering $t_{\rm scatt}$. The multi-beam
survey configured 96 frequency channels over the bandpass, $t_{\rm
  DM}$ can then be represented by that at the central frequency
channel. For $t_{\rm scatt}$, the values given by the NE2001 model
\citep{cl02} were taken. The model gives $t_{\rm scatt}$ at
1,000\,MHz; values at the 1,374\,MHz were extrapolated with the
spectral index $-$4.4 of the Kolmogorov spectrum for turbulence. By
repeating the procedures for zero DM, sensitivities at DM 100, 300 and
1,000\,cm$^{-3}$\,pc were calculated as shown in
Fig. \ref{fig:flux}.

In the figure, the sensitivities derived via the routine which is
originally developed by \citet{cra00} for the multi-beam survey
sensitivity estimates are also shown. We see there are wide
discrepancies between these values and those derived in the
example. This is primarily because the estimates given in the example
were drawn from the 3\,$\sigma$ confidence limit while the original
estimates were drawn from the 8\,$\sigma$ confidence limit. The
effects of the high-pass filters with characteristic times $\sim$2\,s
and the 5\,s cut-off considered in the original estimates would have
further widened the discrepancies at the long period end. The effects
of the filters, cut-off and any other factor that degrade the
sensitivity predictions were not included in the
example. PSR~J1822$-$0848 with period $\sim$2.5\,s and
PSR~J1830$-$0052 with period $\sim$0.3\,s were initially discovered by
the multi-beam survey. The ATNF Pulsar
Catalogue\footnote{http://www.atnf.csiro.au/research/pulsar/psrcat/;
  version 1.56} \citep{mhth05} shows both the pulsars have exhibited a
mean flux density of $\sim$0.04\,mJy. Around the periods of these
pulsars, the sensitivity predictions given in this example are
$\sim$0.01\,mJy, the 3\,$\sigma$ confidence limit is comparatively
low. Around the periods, the original predictions were
$\sim$0.14\,mJy, the 8\,$\sigma$ confidence limit consequently seems
high.
\section*{Acknowledgements}
This work is supported by 1) the National Key Projects of China,
Frontier research on radio astronomy technology; 2) the Open Project
Program of the Key Laboratory of FAST, National Astronomical
Observatories, Chinese Academy of Sciences; 3) the National Natural
Science Foundation of China (No. 11503034, No. 11261140641,
No. 11403060). MY acknowledges Prof. F. Crawford for the helpful
discussion and, in particular, the provision of the original program
for sensitivity calculation of the multi-beam survey. MY acknowledges
Prof. K. J. Lee for the initial discussion. MY also acknowledges the
reviewer for the comments which have greatly helped in improving the
manuscript.
\newpage

%
\newpage
\begin{table}
\vspace{9cm}
\centering
\caption{Optimum numbers of harmonics summed for specific pulse duty
  cycle $\delta$ intervals.}\label{tab:nhrm}
\begin{tabular}{rlcllcl}
\hline \hline \multicolumn{1}{r}{Harm. no.} &
\multicolumn{3}{l}{$\delta$ int. (TopHat)} &
\multicolumn{3}{l}{$\delta$ int. (MVMD)} \\ \hline 1 & 0.55 & -- &
0.44 & 0.50 & -- & 0.39 \\ 4 & 0.44 & -- & 0.38 & & -- & \\ 2 & 0.38 &
-- & 0.24 & 0.39 & -- & 0.22 \\ 4 & 0.24 & -- & 0.12 & 0.22 & -- &
0.10 \\ 8 & 0.12 & -- & 0.061 & 0.10 & -- & 0.050 \\ 16 & 0.061 & -- &
0.029 & 0.050 & -- & 0.025 \\ 32 & 0.029 & -- & 0.0050 & 0.025 & -- &
0.0050 \\ \hline
\end{tabular}  
\end{table}
\newpage 
\newpage
\newpage
\end{document}